\documentclass[preprint,12pt]{elsarticle}

\usepackage{amssymb}

\usepackage{todonotes}

\usepackage{amsmath} 

\usepackage{enumitem}
\usepackage{nicefrac}

\usepackage{booktabs}
\usepackage{makecell}
 
\begin{document}
\begin{frontmatter}



\title{On the inextensibility assumption in the stability of elastic rings: overhaul of a traditional paradigm}

\author[inst1]{Federico Guarracino\corref{cor1}}%
\cortext[cor1]{fguarrac@unina.it, ida.mascolo@unina.it}
\author[inst1]{Ida Mascolo\corref{cor1}}%
\affiliation[inst1]{organization={Department of Structural Engineering, University of Naples "Federico II"},
            city={Naples},
            country={Italy}}

\begin{abstract}
One of the oldest and most common structural engineering issues is the elastic buckling of circular rings under external pressure, which has a fundamental importance in a number of applications in general mechanics, engineering and bio-physics, just to name a few. Levy is considered to have provided the first significant solution to this problem in 1884, and most stability textbooks make reference to this original solution, which is based on the Euler-Bernoulli beam model. Following this incipit, over the past one hundred and forty years a huge number of papers have continued to analyse many special cases and extensions. However, the majority of these studies tend to build on the a-priori assumption of inextensibility of the ring centre line without investigating the real significance and extent of this condition.
Here, in the framework of a suitable non-linear kinematic, the problem is re-examined from its roots, and it is shown that not only the inextensibility paradigm cannot straightforwardly lead to the classic solution in an energy framework, but, on the contrary, the extensibility of the ring is necessary to allow a unified and meaningful treatment of buckling and initial post-buckling behaviour for a complete variety of cases. On these bases, some facts and results in literature are rectified and discussed.
\end{abstract}


\end{frontmatter}

\section{Introduction}
The pre-buckling geometry of a compressed ring can have a much greater influence on its buckling and post-buckling behaviour with respect to what happens in the case of an initially ideal straight column. In fact, differently from Euler's strut, not accounting for the axial deformability of the rod is a much more subtle matter and may lead at least to some misunderstandings if not even misleading results \cite{mascolo2023revisitation}.

The scope of the present article is primarily to overhaul the generally accepted paradigm of axial inextensibility and to present a solution that instead fully accounts for the pre-buckling extensibility of the middle line of the ring. In this manner, it is shown that from an energy standpoint it is possible to point out some hidden issues in all those formulations that assume, on the contrary, an a-priori axial inextensibility of the annular rod.
The proposed solution is the outcome of a rigorous methodological rearrangement of a number of scattered observations over the years, often provided with different goals.

Secondarily, but not less importantly, the presented formulation aims to allow a physically meaningful treatment of buckling and initial post-buckling of elastic rings for a number of instances of practical interest which allows to rectify some known results and to highlight the features of some types of applied loads.

The starting point of the present study has been an examination, to the best of the present authors' capabilities, of the path that seems to constitute the backbone of the problem in its first century of life. 

The plan of the paper is the following: first, the problem is stated in full generality according to Levy's original formulation (Section~\ref{Sec: 1}). 
Then, starting from an energy approach, the equations of equilibrium and their solution are found for a variety of loads (Section~\ref{Sec: 2}). In this respect, the quasi-inextensibility condition yielded by the Castigliano-Menabrea principle is discussed in detail. 
A unified treatment of buckling and initial post-buckling of the system is then presented (Section~\ref{Sec: 3}), highlighting the influence of the axial deformation and amending some results in literature. Finally, some considerations are drawn on the behaviour of the different types of applied loads (Section~\ref{Appendix D}).  
The implications deriving from an a-priori assumption, on the contrary, of the axial inextensibility of the annular rod are  pointed out and illustrated in detail by means of some remarks on the underlying kinematics~(\ref{Appendix C}). A review of the classical and most notable contributions to the topic is finally reported, for the convenience of the reader~~(\ref{Appendix A}).

\section{The general framework}\label{Sec: 1}
\subsection{Basic definitions}\label{Sec: 2-1}
\begin{figure}
    \centering
    \includegraphics[width=\linewidth]{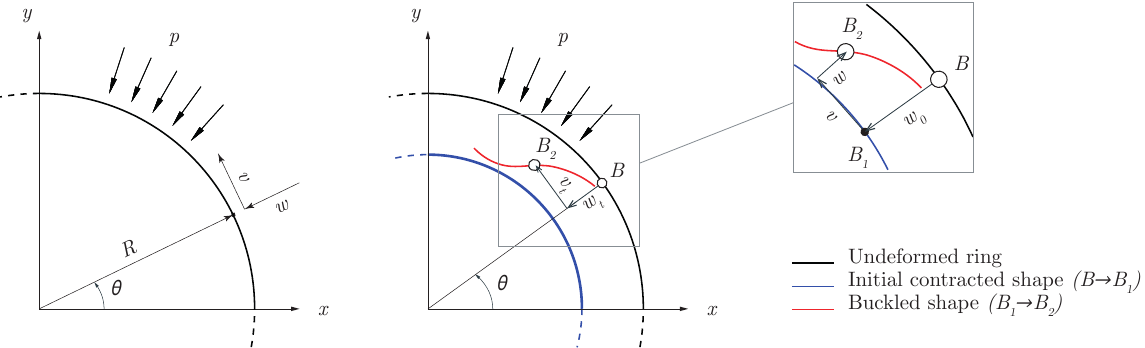}
    \caption{A circular ring under external pressure, from left to right:  initial configuration, uniform contracted status and buckled status. The radial component of displacement is taken as positive when inward; the tangential component is taken as positive when directed as the polar angle \(\theta\) (i.e., anticlockwise).}
    \label{fig:Fig_1}
\end{figure}

The problem is that of the initial buckling of a thin, linearly elastic annular rod with a constant cross section \(S\) and a radius $R$ subject to an external inward pressure $p$ (see Figure~\ref{fig:Fig_1}). The hypothesis are those of the Euler--Bernoulli beam with the shear deformation considered neglectable.

Initially, the middle line of the ring experiences an uniform contraction, $\varepsilon_0$, and its radius decreases evenly. As a result, the ring shrinks while maintaining its original shape (continuous blue line in Figure~\ref{fig:Fig_1}).  
Aside from the shrinking on account of the constant axial force $N_0=-p R$, for a critical value of the external pressure, $p_k$, an asymmetrical, adjacent or buckled status, can take place (continuous red line in Figure~\ref{fig:Fig_1}).

The expressions of the axial stretch $\varepsilon$ (taken positive when it causes ring elongation), of the curvature, $\chi$ of the ring axis and of the radial and tangential components of the displacement, $w$ and $v$, respectively (see Figure~\ref{fig:Fig_1}), can be written as
\begin{subequations}  
\begin{align}
\varepsilon_t &= \varepsilon_0 + \varepsilon=\frac{N_0}{EA}+\varepsilon\,, \label{Eq_0a}\\
\chi_t&= \chi_0 + \chi=\frac{1}{R}-\frac{1}{R+w_0}+\chi\approx\frac{w_0}{R^2}+\chi\,, \label{Eq_0b}\\
w_t&=w_0+w=R \frac{N_0}{EA} +w\,,\label{Eq_0c}\\
v_t&=v_0+v=0+v \,,
\label{Eq_0d}
\end{align}
\end{subequations}
%
\noindent where the subscript \(\left({t} \right)\)\ stands for any generic status and the the subscript \(\left({0} \right)\) stands for the initial uniformly contracted status. 
$EA$ is the axial stiffness of the ring and the pre-buckling deformation is assumed infinitesimal.

The  axial stretch $\varepsilon$,  the rotation of the section of the annular rod $\varphi$ and curvature $\chi$ can be related to the components of displacement by the following relationships \cite{mascolo2023revisitation,babilio2023static}
\begin{subequations}  
\begin{align}
\varepsilon &=\frac{w+v^{\prime}}{R}+\frac{\left(w+v^{\prime}\right)^2+\left(v-w^{\prime}\right)^2}{2 R^2}+\dots\,, \label{Eq_1a}\\
\varphi&=  \arcsin{\left(\frac{v-w^{\prime}}{R}\right)}\, ,\label{Eq_1b}\\
\chi&=\frac{1}{R\;}\varphi^\prime=\frac{v^{\prime}-w^{\prime \prime}}{R^2}\left[1+\frac{1}{2}\left(\frac{v-w^{\prime}}{R}\right)^2+\dots\right] \,,\label{Eq_1c}
\end{align}
\end{subequations}
\noindent where the superscript \(\left(^{\prime} \right)\) indicates differentiation with respect to the angular coordinate $\theta$. 
Retaining only the second-order terms involving $w$, $v$, $w^{\prime}$ and $v^{\prime}$ it is possible simplify the expressions~\eqref{Eq_1c} by neglecting higher-order terms, i.e.,
\begin{subequations}  
\begin{align}
\varepsilon &=\frac{v^{\prime}+w}{R}+\frac{\left(v^{\prime}+w\right)^2+\left(v-w^{\prime}\right)^2}{2 R^2}\,, \label{Eq_1bisa}\\
\varphi&=  \frac{v-w^{\prime}}{R}\, ,\label{Eq_1visb}\\
\chi&=\frac{v^{\prime}-w^{\prime \prime}}{R^2} \,.\label{Eq_1bisc}
\end{align}
\end{subequations}
\subsection{Energy approach}
The total potential energy of the system is given, as usual, by the difference between the strain energy, $U$, and the work done by the applied forces, \(W\), that is
\begin{equation}
    \label{EPT_1}
    \Pi=U-W\,,
\end{equation}
The strain energy can be written as
\begin{align}
   U&=\frac{\text{EA}}{2}\int_{0}^{2\pi} {\varepsilon_t}^2 \,  ds +\frac{\text{EI}}{2} \int_{0}^{2\pi}  {\chi}^2 ds= \nonumber \\
   &=\frac{\text{EA}}{2}\int_{0}^{2\pi} \left({\varepsilon_0}^2+2\,{\varepsilon_0}{\varepsilon}+{\varepsilon}^2\right) \,  ds\,+\frac{\text{EI}}{2} \int_{0}^{2\pi}  {\chi}^2 ds\,,
   \label{StrainEnergy1}
\end{align}
which simplifies retaining also in this case only the terms up to the second order
\begin{align}
   U&=\frac{\text{EA}}{2}\int_{0}^{2\pi} \left({\varepsilon_0}^2+2\,{\varepsilon_0}{\varepsilon}+{\varepsilon}_{lin}^2\right) \,  ds\, +\frac{\text{EI}}{2} \int_{0}^{2\pi}  {\chi}^2 ds\,,
   \label{StrainEnergy}
\end{align}
\noindent where $EI$ is the  flexural stiffness of the annular rod, $ds=Rd\theta$ and \({\varepsilon}_{lin}=\nicefrac{\left({v^{\prime}+w}\right)}{R}\) is the linear part of the deformation \(\varepsilon\).
Equation \eqref{StrainEnergy} can be conveniently rearranged in the following form
\begin{equation}
U=U_0+U_1+U_{0,1}\nonumber \;,
\end{equation}
\noindent where $U_0$ is the strain energy associated with the initial uniform contraction $\varepsilon_0$, $U_1$ is the strain energy associated with the initial buckling shape, and $U_{0,1}$ is the interaction term, that is
\begin{subequations}
    \begin{align}
         U_0&=\frac{\text{EA}}{2}\int_{0}^{2\pi} 
         \varepsilon_0^2\, R d\theta \, ,\label{eq_3a}\\
         U_1&=\frac{\text{EA}}{2}\int_{0}^{2\pi} {{\varepsilon}_{lin}}^2 \, R d\theta +\frac{\text{EI}}{2}\int_{0}^{2\pi} \chi^2 \, R d\theta \, , \label{eq_3b}\\
         U_{0,1}&=\frac{\text{EA}}{2}\int_{0}^{2\pi} 
         2\,\varepsilon_0{\varepsilon} \, R d\theta\, . \label{eq_3c}
    \end{align}
\end{subequations}
The work done by the external load during the initial uniform contraction, that is
\begin{equation}\label{eq_7N}
    {W}_0=\int _{0}^{2\pi}\;  -p\,w_0\, R d\theta= \frac{2\pi\,p^2\, R^3}{EA}\, ,
\end{equation} 
\noindent only depends on the radial component of displacement of the system, and it is not affected by the particular loading type.

On the contrary, when the ring attains the buckled configuration, the work done by the applied load involves additional terms that depend on the type of loading under consideration. 

The types of loading that are usually considered for the problem at hand are 
\begin{itemize}
\item \emph{dead load} (\(d.\))

the load remains normal to the undeflected axis of the element with a constant magnitude per unit initial length;
\item \emph{hydrostatic load} (\(h.\))

the load remains normal to the deflected axis of the element with a constant magnitude per unit initial length. It is the case of the pressure exerted by liquids or gases;
\item \emph{centrally directed load} (\(c.\)) 

the load remains directed towards the centre of the ring with a constant magnitude per unit initial length.  It is the case, for example, of a wheel subjected to many thin spokes under tension;
\item  \emph{inverse-square central load} (\(is.\))

the load remains directed towards the centre of the ring and its magnitude obeys to an inverse-square law with respect to the distance from the centre.
\end{itemize}

\begin{figure}
    \centering
    \includegraphics[width=\linewidth]{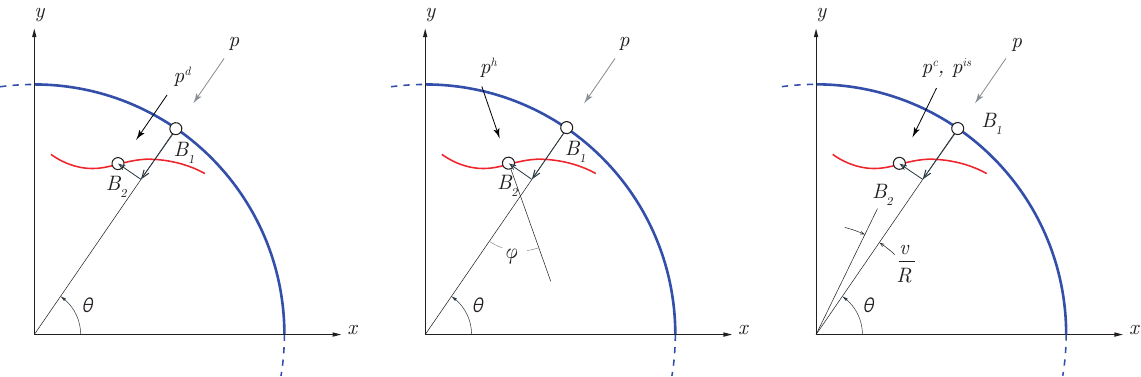}
    \caption{Types of loading: hydrostatic load, $p^{h}$, dead load, $p^{d}$, centrally directed load, $p^{c}$ and inverse-square central load,  $p^{is}$. The solid black and red lines represent the initially contracted shape and the buckled shape, respectively.}
    \label{fig:Fig_2}
\end{figure}

A schematic representation of the different loads behaviours is given in Figure~\ref{fig:Fig_2}. 
It is worth noticing that all types of loading taken into account can be considered conservative~\cite{bodner1958conservativeness} under certain conditions, which will be discussed more in detail in the following. In fact, as pointed out by Oran~\cite{oran1967complementary} certain degree of misconception might be found in the literature, likely fuelled by the fact that the conservative character of loads initially normal to the surface, although conditional, is often taken for granted. For such a reason the work of the different types of load will be here explicitly defined and some consequent implications will be highlighted with respect to the eventual occurrence of rigid motions in Section~\ref{Appendix D}. 

The work done on the ring by the dead load ($d.$) is that of a force that is constant in direction and magnitude per unit initial length and depends only on the initial and final configurations in the absence of any rigid rotation. The assumption that in the case of dead, centrally directed and inverse-square central loads the force is applied per unit initial length is taken here in accordance with the literature~\cite{smith1969effect, budiansky1974theory,gaibotti2024effects} and follows from the fact that these loads are generally conceived as produced by a very large number of closely spaced radial cables attached to the undeformed configuration. The work done by the hydrostatic load ($h.$) is that of a pressure of a liquid at rest, which in the case at hand is obviously a conservative force \cite{Bolotin1963elasticstability}. Finally, the work done by the centrally directed load ($c.$ and $is.$), it is the product of the force by the change in the distance from the pole, so that work results independent of the path in the absence of any rigid translation. 

The work of the applied load can be written as follows
\begin{equation}
  W^{i}=W_0+W_1^{i}\,, \qquad {i}= d,\,h,\, c, \, is,\,
\end{equation}
where the superscript $i$ states for the different loading cases.

In particular, the contribution of the work of the hydrostatic load, \(W^h_1\), at buckling can be expressed in terms of the product of the pressure \(p\) times the change in the area enclosed by the centroidal surface of the ring, which can be exactly calculated by means of the following formula, as shown in  \cite{mascolo2023revisitation}

\begin{equation}
    W_1^{h}= -\frac{p}{2}\int _{0}^{2\pi}w\left(w+v^{\prime}\right)-\;v \left(v-w^{\prime}\right) \,d\theta        \,.
\label{Eq_AppB}
\end{equation}

In the loading case \(d.\), the elementary force $p^{d} \, ds=p Rd\theta$ (Figure~\ref{fig:Fig_2}) does not vary in amplitude per unit initial length or direction at the onset of buckling, thus it only involves the radial displacement component \(w\) and the contribution of the work of the force can be exactly written as
\begin{equation} 
        W_1^{d}=-\frac{p}{2}\int _{0}^{2\pi}\;2R\,w\,d\theta\,.
\label{Eq_9tot}
\end{equation}

In both cases, the work \(W_1^i\) is associated with the variation in the radius length so that, by integrating over the displacement path it is 
\begin{subequations} 
\begin{align}
        W_1^{c}=& -p\int_{0}^{2\pi}\; \left(R -\sqrt{(R + w)^2+v^2}\right)\,Rd{\theta}=\nonumber\\
        =&W_1^{d}- \frac{p}{2}\int _{0}^{2\pi}\;\left[v^2 \left(1-\frac{w}{R}+\frac{w^2}{R^2}-\frac{v^2}{4 R^2}\right)+\dots\right]\,d\theta \;,\label{Eq_9cent}\\
        W_1^{is}=& -p\int _{0}^{2\pi}\; \left(-{R^2}+\frac{R^3}{\sqrt{v^2+(R+w)^2}}\right)\,Rd{\theta}=\nonumber\\
        =&W_1^{d}- \frac{p}{2}\int _{0}^{2\pi}\;\left[v^2 \left(1-\frac{3 w}{R}+\frac{6 w^2}{R^2}-\frac{3 v^2}{4 R^2}\right)+\right.\nonumber\\&\;\left.+2 w^2 \left(-1+\frac{w}{R}-\frac{w^2}{R^2}\right)+\dots\right]\,d\theta \;.\label{Eq_9is}
\end{align}\label{Eq_94ord}
\end{subequations}

It is worth noticing that equations~\eqref{Eq_94ord} coincide with the four-order approximation of the formulae reported by Budiansky~\cite{budiansky1974theory,sills1978postbuckling}. 

\noindent Overall, up to the second order of magnitude, it is
\begin{subequations} 
\begin{align}
        W_1^{d}&=-\frac{p}{2}\int _{0}^{2\pi}\;2R\,w\,d\theta\,,\label{Eq_9cF}\\
        W_1^{h}&=W_1^{d}- \frac{p}{2}\int _{0}^{2\pi}\,v \left(v-w^{\prime}\right) \,d\theta\,, \label{Eq_9bF}\\
        W_1^{c}&=W_1^{d}- \frac{p}{2}\int _{0}^{2\pi}\;v^2\,d\theta \,,\label{Eq_9dF}\\
        W_1^{is}&=W_1^{d}- \frac{p}{2}\int _{0}^{2\pi}\;v^2-2w^2\,d\theta \;.\label{Eq_9eF}
\end{align}\label{Eq_9totFin}
\end{subequations}
\noindent Thus, the expression of the total potential energy of the system becomes
\begin{subequations}  
\begin{align}
\Pi^{i}&=U-W^{i}\,, \label{Eq_11a} \\ 
\Pi^{d}&=U_0-W_0+\left(\frac{EA}{2 R}-\frac{p}{2}\right) \int_{0}^{2\pi} \left(w+v^{\prime}\right)^2 \, d\theta\, + \nonumber\\&+\frac{EI}{2 R^3}\int_{0}^{2\pi} \left(v^{\prime}-w^{\prime \prime}\right)^2 \, d\theta-\frac{p}{2} \int_{0}^{2\pi} \left(v-w^{\prime}\right)^2 \, d\theta \;,  \label{Eq_11c}\\
\Pi^{h}&=\Pi^{d}+\frac{p}{2} \int_{0}^{2\pi} v \left(v-w^{\prime}\right) \, d\theta \;,  \label{Eq_11b} \\
\Pi^{c}&=\Pi^{d}+\frac{p}{2} \int_{0}^{2\pi} v^2 \, d\theta \;, \label{Eq_11d}\\
\Pi^{is}&=\Pi^{d}+\frac{p}{2} \int_{0}^{2\pi} v^2 - 2w^2 \, d\theta \;. \label{Eq_11e}
\end{align}\label{Eq_11tot}
\end{subequations}

By introducing the radius of gyration $\rho=\sqrt{\nicefrac{I}{A}}$, $H=(\nicefrac{\rho}{R})^2$ , $\lambda =\nicefrac{R^3 p}{\text{EI}}$, \(\kappa=\nicefrac{2R^3}{EI}\) and \(\tilde{\Pi}_0=\kappa \left(U_0+W_0\right)=const\), the total potential energy of the system can be written in a dimensionless form \(\tilde{\Pi}=\kappa\, \Pi\) as follows
\begin{subequations}  
\begin{align}
\tilde{\Pi}^{d}&=\tilde{\Pi}_0+\frac{1-H \lambda}{H} \int_{0}^{2\pi} \left(w+v^{\prime}\right)^2 \, d\theta + \int_{0}^{2\pi} \left(v^{\prime}-w^{\prime \prime}\right)^2 \, d\theta \nonumber\\&  -\, \lambda \int_{0}^{2\pi} \left(v-w^{\prime}\right)^2 \, d\theta  \;,\label{Eq_12bb}\\
\,\tilde{\Pi}^{h}&=
\,\tilde{\Pi}^{d}+\, {\lambda} \int_{0}^{2\pi} v \left(v-w^{\prime}\right) \, d\theta \;,  \label{Eq_12aa} \\
\tilde{\Pi}^{c}&=\,\tilde{\Pi}^{d}+\, {\lambda} \int_{0}^{2\pi} v^2 \, d\theta \;, \label{Eq_12cc}\\
\tilde{\Pi}^{is}&=\tilde{\Pi}^{d}+\, {\lambda} \int_{0}^{2\pi} v^2-2w^2 \, d\theta \;. \label{Eq_12dd}
\end{align}
\end{subequations} 
\subsection{Euler-Lagrange equations}
The Euler-Lagrange equations associated with the stationary of the energy functional \(\overline{\Pi}^i\) result, in full generality, the following
\begin{subequations}  
\begin{align}
\textit{d.} \quad
& H w^{(3)}-(1+ H-H\lambda) v^{\prime \prime}-(1-2 H \lambda)\, w^{\prime}-H \lambda \, v=0 \nonumber\\
&H \left(\lambda  w^{\prime \prime}-v^{(3)}+w^{(4)}\right)+(1-2 H \lambda ) v^{\prime}+(1-H \lambda) w=0
 \label{Eq_N25b}\\
\nonumber \\ 
\textit{h.} \quad
& H w^{(3)}-\left(1+ H-H\lambda\right) v^{\prime \prime}-\left({1}-{\frac{3}{2}H}\lambda \right)w^{\prime}=0\nonumber \\
&H \left(\lambda  w^{\prime \prime}-v^{(3)}+w^{(4)} \right)+\left({1}-{\frac{3}{2}H}\lambda \right) v' +(1-H \lambda)w =0
\label{Eq_N25a} \\ 
\nonumber \\
\textit{c.} \quad
&  H w^{(3)}-(1+ H-H\lambda)v^{\prime \prime}-(1-2 H \lambda ) w^{\prime}=0\nonumber\\
&H \left(\lambda  w^{\prime \prime}-v^{(3)}+w^{(4)}\right)+(1-2 H \lambda ) v^{\prime}+(1-H \lambda)w=0\,\\ 
\nonumber \\ 
\textit{is.} \quad
& H w^{(3)}-(1+ H-H\lambda)v^{\prime \prime}-(1-2 H \lambda ) w^{\prime}=0\nonumber\\
&H \left(\lambda  w^{\prime \prime}-v^{(3)}+w^{(4)}\right)+(1-2 H \lambda ) v^{\prime}+(1-3H \lambda)w=0\,.
\label{Eq_N25c}
\end{align}
\end{subequations}
These equations can be written in compact form \cite{DYM1971, simitses2006fundamentals} as
\begin{equation}
\bigg \{\begin{array}{rl}
L_1^{i} w+{L_2^{i}} v=0 \\
L_1^{i} v+{L_3^{i}} w=0 \label{Eq_26}
\end{array}
\end{equation} 
\noindent where
\begin{subequations}
\begin{align}
L_1^{d}&=H f^{(3)}-\left({1-2 H \lambda }\right) f^{\prime} \nonumber\\
L_2^{d}&=(1+H- H \lambda) f^{\prime \prime}+H \lambda f\\
L_3^{d}&=H f^{(4)}+H \lambda  f^{\prime \prime}+(1-H \lambda)f \nonumber
\label{Eq_26b}
\end{align}
\begin{align}
L_1^{h}&=H f^{(3)}-\left(1-\frac{3}{2} H \lambda\right) f^{\prime}\nonumber\\
L_2^{h}&=(1+H- H \lambda) f^{\prime \prime}\\
L_3^{h}&=H f^{(4)}+H \lambda  f^{\prime \prime}+(1-H \lambda)f \nonumber
\label{Eq_26a}
\end{align}
\begin{align}
L_1^{c}&=H f^{(3)}-\left(1-\frac{1}{2} H \lambda\right) f^{\prime}\nonumber\\
L_2^{c}&=(1+H- H \lambda) f^{\prime \prime}\\
L_3^{c}&=H f^{(4)}+H \lambda  f^{\prime \prime}+(1-H \lambda)f \nonumber
\label{Eq_26c}
\end{align}
\begin{align}
L_1^{is}&=H f^{(3)}-\left(1-{2} H \lambda\right) f^{\prime}\nonumber\\
L_2^{is}&=(1+H- H \lambda) f^{\prime \prime}\\
L_3^{is}&=H f^{(4)}+H \lambda  f^{\prime \prime}+\left(1-3 H\lambda \right)f \nonumber
\label{Eq_26d}
\end{align}
\end{subequations}

\noindent Manipulating equations \eqref{Eq_26} it is possible to write
\begin{equation}
\bigg \{
\begin{array}{c}
{L_1^{i}} {L_3^{i}} \,w+{L_2^{i}} {L_3^{i}} \,v=0 \\
{L_1^{i}} {L_3^{i}} \,w+{L_1^{i}} {L_1^{i}} \,v=0 
\end{array}\label{Eq_27}
\end{equation}
\noindent and by a simple addition
\begin{equation}
\left({L_2^{i}} {L_3^{i}}-{L_1^{i}}^2\right) v + \left({L_1^{i}} {L_3^{i}}-{L_1^{i}} {L_3^{i}}\right) w=0\,. \label{Eq_28}
\end{equation}
\noindent Since $({L_1^{i}} {L_3^{i}}-{L_1^{i}} {L_3}^{i})=0$, it follows
\begin{equation}
\left({L_2^{i}} {L_3^{i}}-{L_1^{i}}^2\right) v=0\,,\label{Eq_29}
\end{equation}
\noindent which involves the tangential component of displacement, \(v\), only.

\noindent For each of the different loading cases, it is
\begin{subequations}\label{Eq_32tot}
\begin{align}
\textit{d.}  \quad
&v^{(6)}+\left(2+\lambda \right)v^{(4)}+\left(1+2 \lambda \right) v^{\prime \prime}\nonumber+{ \lambda  }v=0\;,\\
\textit{h.}  \quad
&v^{(6)}+ \left(2 + \lambda\right)v^{(4)}+\left(\frac{1}{1-H \lambda }+\lambda \right)v^{\prime \prime}=0\;,
\label{Eq_32a}\\
\textit{c.} \quad
&v^{(6)} +\left(\frac{2}{1-H \lambda }+\lambda\right) v^{(4)}+\left[\frac{1}{1-H \lambda }+\right.\nonumber\\&\left.+2 \lambda  \left(1-\frac{9 H \lambda }{8 (1-H \lambda )}\right)\right] v^{\prime \prime} =0\;,\\ 
\textit{is.} \quad
&v^{(6)} +\left(3-\frac{1}{1-H \lambda }+\lambda \right) v^{(4)}+ \left[3 \left(1-\frac{2}{3 (1-H \lambda )}\right)+\right.\nonumber\\&\left.+\lambda  \left(1-\frac{1}{1-H \lambda }\right)\right] v^{\prime \prime}=0 \;. 
\end{align}\label{Eq_N32}
\end{subequations}

Finally, assuming that the axial deformation of the ring, i.e., \( H\lambda=\nicefrac{R \, p}{EA}\), can be neglected with respect to the unity at the onset of buckling and only at this stage, it is possible to simplify equations \eqref{Eq_N32} as follows  
\begin{subequations}  
\begin{align} 
\textit{d.} \qquad &v^{(6)}+(2+\lambda)\,v^{(4)}+(1+2 \lambda)\, v^{\prime \prime}+ \lambda \, v=0 \,,\label{Eq_18b}\\
\textit{h.} \qquad &v^{(6)}+(2+\lambda)\,v^{(4)}+(1+ \lambda)\, v^{\prime \prime}=0 \,,\label{Eq_18a}\\
\textit{c.} \qquad &v^{(6)}+(2+\lambda)\,v^{(4)}+(1+2 \lambda)\, v^{\prime \prime}=0\,,\label{Eq_18c}
\\
\textit{is.} \qquad &v^{(6)}+(2+\lambda)\,v^{(4)}+ v^{\prime \prime}=0\,.\label{Eq_18d}
\end{align}
\label{Eq_18tot}
\end{subequations}
\subsection{The role of the principle of Castigliano--Menabrea}

It is here worth pointing out that the assumption \(H\lambda<<1\) has an energy connection and a consequent kinematic implication. In fact, the condition for the existence of two adjacent equilibrium states of the ring, the circular one and a slightly non-circular one, under the same load intensity is that the first variation of the potential energy with respect to a small displacement of the ring from its circular status is zero. 

Now, by virtue of the principle of minimum strain energy by Castigliano--Menabrea \cite{southwell1936castigliano} only the displacement field that minimizes the dimensionless functional
\begin{align}   
\tilde{U}=&\kappa\, U_0+\frac{{1}-\lambda{H}}{H}\int_{0}^{2\pi} \left(w+v^{\prime}\right)^2 \, d\theta +  \int_{0}^{2\pi} \left(v^{\prime}-w^{\prime \prime}\right)^2 \, d\theta + \nonumber
\\ &-\lambda \, \int_{0}^{2\pi} \left( v-w^{\prime}\right)^2  \, d\theta\,, 
\label{Eq_12}
\end{align}
\noindent are actual states of equilibrium for the ring. Given that the constant term \(\kappa\,U_0\) and the subtractive term are immaterial to the problem, attention can be focused on the sum 
\begin{equation}   
\frac{{1}-\lambda{H}}{H}\int_{0}^{2\pi} \left(w+v^{\prime}\right)^2 \, d\theta +  \int_{0}^{2\pi} \left(v^{\prime}-w^{\prime \prime}\right)^2 \, d\theta\,.
\label{Eq_12double}
\end{equation}
Therefore, since 
\begin{equation}
   \frac{{1}-\lambda{H}}{H}\approx  \frac{{1}}{H}>>1\,,
   \label{Eq_13}
\end{equation}
\noindent all the displacement fields for which the term 
\begin{equation}
   \int_{0}^{2\pi} \left(w+v^{\prime}\right)^2 \, d\theta \,,\label{Eq_13a}
\end{equation}
\noindent has a value different from zero are not apt to minimize the strain energy  \eqref{Eq_12}.

On the contrary, Chwalla and Kollbrunner~\cite{chwalla1938beitrage} obtained equations~\eqref{Eq_18tot} under the a-priori kinematic condition
\begin{equation}
w = -v^{\prime} \;,
\label{Eq_inex}
\end{equation} 
\noindent which makes the linear part of the axial strain~\eqref{Eq_1a}, $\varepsilon_{lin}$, always null.

Here equations~\eqref{Eq_12}--\eqref{Eq_13a} mark a significant point in the understanding of ring buckling and opens the way to the clarification of a number of hidden inconsistencies found in many classic works and to a straightforward exploration of  the post-buckling behaviour of the system by allowing to take it into account both in the pre-buckling and in the post-buckling the extensional part of the strain energy.

\section{Solution of the buckling problem}
\label{Sec: 2} 

The solution of the differential equations \eqref{Eq_18tot} provides
\begin{subequations}  
\begin{align}  
v^{d}=\,&C_1^{d} \sin{\theta}+C_2^{d} \cos{\theta} +C_3^{d} \theta  \sin{\theta}+C_4^{d} \theta  \cos{\theta}+C_5^{d} \sin\theta  \sqrt{\lambda}+ \nonumber \\
&+C_6^{d} \cos{\theta  \sqrt{\lambda }}\label{Eq_19b}\\
v^h= \,&
\frac{C_1^{h}-C_2^{h} }{\left(\lambda^{h}\right)^2}\sinh \left(\theta  \lambda^{h}\right)+
\frac{C_1^{h} +C_2^{h} }{ \left(  \lambda^{h}\right)^2} \cosh \left(\theta  \lambda^{h}\right) -C_3^{h} \cos (\theta )+\nonumber\\&-C_4^{h} \sin (\theta )+C_5^{h}+C_6^{h} \theta \\
v^{c}=\,&
C_1^{c} \sinh \left(\theta  \lambda _1^{c}\right)+C_2^{c} \sinh \left(\theta  \lambda _2^{c}\right)+C_3^{c} \cosh \left(\theta  \lambda _1^{c}\right)+\nonumber\\
&+C_4 ^{c}\cosh \left(\theta  \lambda _2^{c}\right)
+C_5^{c}+C_6^{c} \theta
\label{Eq_19c}\\
v^{is}=\,&
\frac{C_1^{is}- C_2^{is}}{\left(\lambda _2^{is}\right)^2}\sinh \left(\theta  \lambda _2^{is}\right)+\frac{C_1^{is}+ C_2^{is}}{\left(\lambda _2^{is}\right)^2}\cosh \left(\theta  \lambda _2^{is}\right)+\\
+&\frac{C_3^{is}- C_4^{is}}{\left(\lambda _1^{is}\right)^2}\sinh \left(\theta  \lambda _1^{is}\right)+\frac{C_3^{is}+ C_4^{is}}{\left(\lambda _1^{is}\right)^2}\cosh \left(\theta  \lambda _1^{is}\right)+ C_5^{is}+ C_6^{is}\theta
\label{Eq_19d}
\end{align}
\label{EqN_19}
\end{subequations}

\noindent where
\begin{subequations}  
\begin{align}
    &\lambda^{\textit{h}}= \sqrt{-1-\lambda}\,,
    \label{Eq_19d1}\\
    &\lambda _{1,2}^{\textit{c}}= \sqrt{-\left(1+\frac{\lambda }{2}\right)\pm \sqrt{\lambda  \left(\frac{\lambda }{4}-1\right)}}\,,\\
    &\lambda _{1,2}^{\textit{is}}= \sqrt{-\left(1+\frac{\lambda }{2}\right)\pm \sqrt{\lambda  \left(\frac{\lambda }{4}+1\right)}}\,.\label{Eq_19d2}
\end{align}
\end{subequations}

At this stage, in order to rule out from buckling any rigid translation, all linear combinations of \( \sin{\theta} \) and \(  \cos{\theta} \)
have to be null, as well as any constant term, given that all these latter can be interpreted as an rigid motion, as it will be discussed in Section~\ref{Appendix D}. Moreover, for \(2\pi\)-periodicity any combination of the secular terms $\theta$, $\theta\,\cos{\theta}$ and $\theta\,\sin{\theta}$ must vanish.

Solutions~\eqref{EqN_19} can be thus rewritten as
\begin{subequations}  
\begin{align}  
v^{d}=\,&
C_5^{d} \sin\theta  \sqrt{\lambda}+C_6^{d} \cos{\theta  \sqrt{\lambda }}\,,\label{Eq_19bN}\\
v^h= \,&
\frac{C_1^{h}-C_2^{h} }{\left(\lambda\right)^2}\sinh \left(\theta  \lambda\right)+
\frac{C_1^{h} +C_2^{h} }{ \left(  \lambda\right)^2} \cosh \left(\theta  \lambda\right)\,,\\ 
v^{c}=\,&
C_1^{c} \sinh \left(\theta  \lambda _1\right)+C_2^{c} \sinh \left(\theta  \lambda _2\right)+C_3^{c} \cosh \left(\theta  \lambda _1\right)+\nonumber\\
&+C_4 ^{c}\cosh \left(\theta  \lambda _2\right)\,,
\label{Eq_19cN}\\
v^{is}=\,&
\frac{C_1^{is}- C_2^{is}}{\left(\lambda _2^{is}\right)^2}\sinh \left(\theta  \lambda _2^{is}\right)+\frac{C_1^{is}+ C_2^{is}}{\left(\lambda _2^{is}\right)^2}\cosh \left(\theta  \lambda _2^{is}\right)+\\
+&\frac{C_3^{is}- C_4^{is}}{\left(\lambda _1^{is}\right)^2}\sinh \left(\theta  \lambda _1^{is}\right)+\frac{C_3^{is}+ C_4^{is}}{\left(\lambda _1^{is}\right)^2}\cosh \left(\theta  \lambda _1^{is}\right)\,.
\label{Eq_19dN}
\end{align}
\label{EqN_19N}
\end{subequations}
Actually, general conditions of continuity can be imposed on the solutions \eqref{EqN_19N}, i.e.
\begin{equation}  
v[0]=v[2\pi], \quad v^\prime[0]=v^\prime[2\pi], \quad \dots\,, \quad v^{(n)}[0]=v^{(n)}[2\pi]\,,  
\label{Eq_continuity}
\end{equation}
\noindent and eventually a system of homogeneous equations is obtained. By setting the determinant of the coefficients of these equations equal to zero, a set of characteristic equations is obtained whose roots represent the eigenvalues \(\lambda\) of the system as functions of the natural numbers \( n=0,1,2,\dots\)  
\begin{subequations}  
\begin{align}  
\textit{d.} \quad & 
\lambda= 4n^2
\,, \quad  \lambda= \left(1+2n\right)^2 
\,,\label{Eq_22b}\\
\textit{h.} \quad & 
\lambda= 4 n^2-1 \,, \quad  \lambda= 4n (1+n)\,, 
\label{Eq_22a}\\
\textit{c.} \quad &  
\lambda=\frac{\left(n^2-1\right)^2}{n^2-2}\,, \label{Eq_22c}\\
\textit{is.} \quad &  
\lambda=\frac{\left(n^2-1\right)^2}{n^2}\,.  \label{Eq_22d}
\end{align}\label{Eq_N22}
\end{subequations}

Interestingly, for dead loading, the lowest physically admissible eigenvalue would seem to be \(\lambda=1\), but it corresponds to an unbuckled mode and therefore does not indicates an equilibrium bifurcation point.
Actually, for dead load the lowest significant eigenvalue results \(\lambda = 4\). The lowest physically admissible eigenvalue for the case of hydrostatic loading \eqref{Eq_22a} results, for \(n=1\), \(\lambda=3\). This result coincides, in full generality, with the classical solution by Levy \cite{levy1884memoire}.

In the case of centrally directed loading, the lowest physically significant eigenvalue, for \(n=2\) is \(\lambda=\nicefrac{9}{2}\)
and in the case of inverse-square central load for \(n=2\) is \(\lambda=\nicefrac{9}{4}\).

Consequently, the solutions of the equations \eqref{Eq_18tot} which actually correspond to the lowest significant bifurcation loads in absence of any rigid motion, are 
\begin{equation}  
\begin{split}   
\textit{d.} \quad 
&\overline{\lambda}^d=4\,, \quad v^d=C \cos{2\theta} \, \quad\text{and}\, \quad w^d=2 C \sin{2\theta} \,, \\
\textit{h.} \quad 
&\overline{\lambda}^h=3\,, \quad v^h=C \cos{2\theta} \, \quad\text{and}\, \quad w^h=2 C \sin{2\theta}\,,\\
\textit{c.} \quad  
&\overline{\lambda}^c=\frac{9}{2}\,, \quad v^c=C \cos{2\theta} \, \quad\text{and}\, \quad w^c=2 C \sin{2\theta}\,, \\
\textit{is.} \quad  
&\overline{\lambda}^{is}=\frac{9}{4}\,, \quad v^{is}=C \cos{2\theta} \, \quad\text{and}\, \quad w^{is}=2 C \sin{2\theta}\,.
\end{split}\label{Eq_19bis}
\end{equation}

\section{Analysis of the initial post-buckling behaviour} \label{Sec: 3}
In order to analyse the post-buckling behaviour of the elastic rings under the various loading conditions examined in the previous Sections, resort is here made to a non-linear Rayleigh-Ritz analysis in which the linear buckling mode is employed as the assumed form~\cite{thompson1964eigenvalue}.

In fact, while a classic procedure for examining the initial post-buckling behaviour of a structure developed by Koiter~\cite{koiter1962theory} makes use of the presumably known linear buckling mode, involving the solution of a set of linear equations to determine the small initial changes in the buckled form, a much straightforward procedure, as discussed in details by Thompson~\cite{thompson1964eigenvalue} is to use the known linear buckling mode as the assumed form in a non-linear Rayleigh-Ritz analysis.

Actually, it is assumed that the deformation of the elastic ring can only be analysed into mode-forms, represented by the eigenfunctions~\eqref{Eq_19bis} of the buckling problem, the amplitudes of which will supply the generalized coordinates for the system. 

Of course the method can only be used in the context of an appropriate non-linear energy formulation, which is here provided by equations~\eqref{Eq_11tot}, accounting for both axial and flexural deformability.
In~\ref{Appendix C} it will be shown that this approach cannot, on the contrary, be straightforwardly applied to any a-priori inextensible formulation because in these cases the expression of the work of the applied load must be manipulated under certain assumptions in order to get the correct buckling values of the load~\eqref{Eq_19bis}, a fact that, however, implies hidden kinematic inconsistencies. 

The behaviour of the annular rod is studied in the vicinity of the first critical equilibrium state, by analysing the non-linear solution of the equilibrium equations \(\nicefrac{\partial \Pi_i}{\partial C}\), which can be considered correct in the immediate vicinity of this state. As stated above, the amplitude \(C\) of the critical eigenfunctions~\eqref{Eq_19bis} is the only free parameter of the analysis. 

Reference is made to the expression of the total potential energy of the system approximated up to the fourth order by accounting for equations~\eqref{Eq_1a} and~\eqref{Eq_1c}  in the expression of the strain energy~\eqref{StrainEnergy1} and series expansions of equations for the work of applied pressure~\eqref{Eq_94ord}.  

It is worth noticing that in the case of the deal load, being the work of the external pressure null, at buckling the phenomenon is triggered by the elastic energy stored in the ring as a consequence of the pre-buckling work of the applied pressure.

Use of critical eigenfunctions~\eqref{Eq_19bis} leads to write Table~\ref{Tab_3.1} and the equilibrium condition \(\nicefrac{\partial \Pi_i}{\partial C}\) provides the non-linear relationship between the applied pressure and the amplitude \(C\) of the critical eigenfunction assumed as the Rayleigh-Ritz shape function, see Table~\ref{Tab_3.2}.

\begin{table}[ht!]
\caption{Present extensible formulation \eqref{Eq_EPT ext} with \(v[\theta]=C\, \cos{2\theta}\) and \(w[\theta]=2C\, \sin{2\theta}\)}
  \centering
  \begin{tabular}{cccccc} 
  \hline
  Load&$U_{0}$&$W_{0}$&$U_{1}$ &$U_{0,1}$ &  $W_1$\\ 
    \hline 
    \makecell[c]{\textit{d}. \\\textit{h}. \\ \textit{c}.\\ \textit{is}.}
    &\makecell[c]{\scriptsize{$\frac{\pi \,p^2 R^3}{EA}$}} 
    &\makecell[c]{\scriptsize{$\frac{2 \pi R^3p^2}{EA}$}} 
    &\makecell[c]{\scriptsize{$\frac{243 \pi  \text{EA} }{32 R^3}\left(C^4+\right.$}\\ \scriptsize{$\left.+\frac{16}{3} H \left(C^4+\frac{4 R^2 C^2}{9 }\right)\right) $}} 
    &\makecell[c]{\scriptsize{$-\frac{9 \pi}{2} \,C^2  p$}} 
    & \makecell[c]{\scriptsize{$0$}\\ \scriptsize{$\frac{3 \pi}{2} \,C^2 p$}\\ \scriptsize{$-\frac{\pi\, }{2} \left( \,C^2  +\frac{13}{16 R^2} C^4 \right)p$}\\ \scriptsize{$\frac{\pi\, }{2} \left( 7\,C^2 +\frac{297}{16 R^2} C^4  \right)p$}}
    \\ 
    \bottomrule
  \end{tabular}\label{Tab_3.1}
\end{table}
\begin{table}[ht!]
\caption{Extensible formulation with \( v[\theta] = C \cos{2\theta} \) and \( w[\theta] = 2C \sin{2\theta} \)}
\centering 
\begin{tabular}{ccc} 
\toprule
Load type & \( \lambda \) & \( \overline{\lambda} \) \\ 
\midrule
\textit{d}. & \( 4 + \frac{27}{16} \left(1+\frac{16}{3} H\right) \frac{C^2}{\rho^2}\) & \( 4 \) \\
\textit{h}. & \( 3+\frac{81}{64}\left({1+\frac{16 }{3}H}\right)\frac{C^2}{\rho^2} \) & \( 3 \) \\
\textit{c}. & \( \left(\frac{9}{2}+\frac{243}{128}\left(1+\frac{16 }{3}H\right)\right)\left(1-\frac{13 C^2}{128 R^2}\right)^{-1}\frac{C^2}{\rho^2}  \)  & \( \nicefrac{9}{2} \)\\
\textit{is}. & \( \left(\frac{9}{4}+\frac{243}{256}\left(1+\frac{16 }{3}H\right)\right)\left(1+\frac{321 C^2}{256 R^2}\right)^{-1}\frac{C^2}{\rho^2}  \) & \( \nicefrac{9}{4} \) \\
\bottomrule
\end{tabular}
\label{Tab_3.2}
\end{table}

Trivially, the values of the buckling multipliers \(\overline{\lambda}\) in Table~\ref{Tab_3.2} are obtained by taking the limit of the non-linear relationships, \(\lambda=\lambda\left(C\right)\), for \(C\rightarrow0\). 
\begin{figure}
    \centering
    \includegraphics[width=0.6\linewidth]{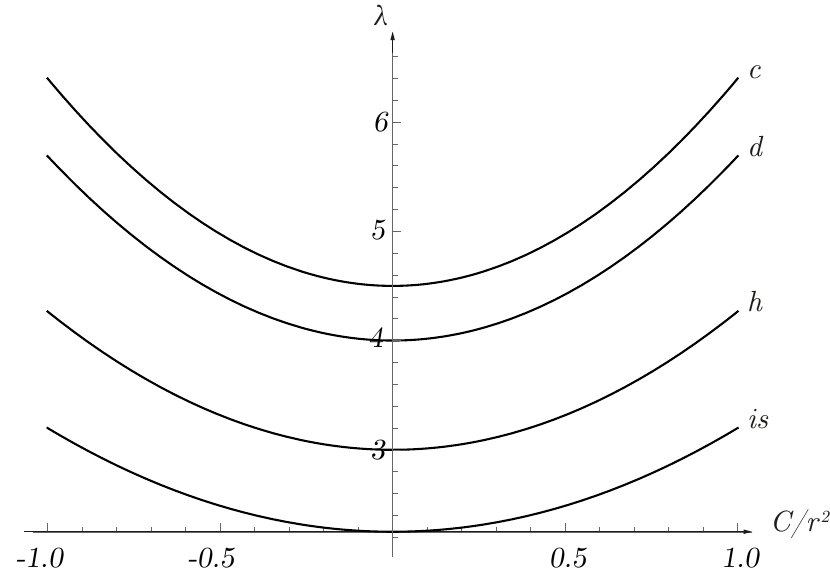}
    \caption{Post-buckling behaviour in the various loading cases.}
    \label{fig:5}
\end{figure}

Beyond the onset of buckling, the non-linear relationships of Table~\ref{Tab_3.2} provide the initial post-buckling behaviour for all the considered load cases and are plotted in Figure~\ref{fig:5}. The initial post-buckling behaviour results stable in all four cases, including the case of inverse square central loading, differently from what was reported in~\cite{sills1978postbuckling} as a consequence of a long lasting debate started by~\cite{el1976influence}. 

It is here important to point out that the non-linear post-buckling behaviour of elastic rings results mostly governed by the terms related to the axial stiffness, \(EA\), as it happens in the von Karmann analysis of the stability of thin elastic plates~\cite{guarracino2007considerations} and it has been observed in the case of cylindrical shells~\cite{1970CrollChilver}. This can be observed by the fact that the contribution of the flexural stiffness \(EI\) to the post-buckling behaviour is represented by the terms in Table~\ref{Tab_3.2} multiplied by the factor \(H\) which is \(<<1\).

\section{The role of the rigid motion}\label{Appendix D}

As shown in Section~\ref{Sec: 2}, the solutions of the equations~\eqref{Eq_18tot} which yield the lowest bifurcation loads in the absence of any rigid motion, are the formulae~\eqref{Eq_19bis}, which in fact have been derived by ruling out any such motion. 
The problem of rigid motion has been recently reviewed in detail, even if from a different standpoint, in~\cite{gaibotti2024effects} where, noticeably, an experimental testing of a ring under central loading has also been provided. 

\begin{figure}
    \centering
    \includegraphics[]{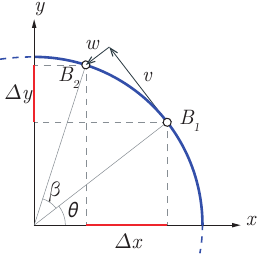}
    \caption{Rigid rotation.}
    \label{fig:enter-label}
\end{figure}
In the energy framework proposed here, one can notice that a rigid-body displacement field for the ring is given, in the vicinity of the undeformed configuration, by the superposition of a horizontal displacement, \(\alpha_1\), and of a vertical displacement, \(\alpha_2\)
\begin{subequations}    \label{Eq_39} 
    \begin{align}
    & v=-\alpha_1 \sin\theta+  \alpha_2 \cos\theta
     \,,\\
    & w=\alpha_1 \cos\theta +  \alpha_2 \sin\theta \, ,  
    \end{align}
\end{subequations}  
plus a rotation about the centre of the ring, \(\beta\), which can be derived from basic considerations of differential geometry (see Figure~\ref{fig:enter-label}) by solving the following system of equations 
\begin{subequations}    \label{Eq_39_1} 
    \begin{align}
    \Delta x=v \sin{\theta}+w \cos{\theta}& =-R \left(\cos{\theta}\cos{\beta} - \sin{\theta}\sin{\beta}-\cos{\theta} \right)\approx \nonumber \\&\approx R\,\left(\frac{\beta^2}{2}\cos{\theta}+\beta \sin{\theta}   \right)+o(\beta^3)\, ,\\
    \Delta y=v \cos{\theta}- w \sin{\theta}& =R \left(\sin{\theta}\cos{\beta} + \cos{\theta}\sin{\beta}-\sin{\theta} \right) \approx \nonumber \\&\approx R\,\left(-\frac{\beta^2}{2}\sin{\theta}+\beta \cos{\theta}   \right)+o(\beta^3)\,
     \, ,   
    \end{align}
\end{subequations}  
which yields~\cite{singer1970buckling}
\begin{subequations}    \label{Eq_39b} 
    \begin{align}
    & v \approx R\,\beta  \, ,\\
    & w\approx R \frac{\beta^2}{2}
     \, .   
    \end{align}
\end{subequations}  

By taking only the linear terms in equations~\eqref{Eq_39} and~\eqref{Eq_39b}, it is immediately seen that the general solutions of differential equations~\eqref{Eq_18tot}, i.e. formulae~\eqref{EqN_19}, show that the buckling pressure for the dead load results neutral with respect to possible rigid rotations, for the hydrostatic load results indifferent to possible rigid rotation and translation, and for the centrally directed loads (\(c.\) and \(is.\)) it is indifferent to any possible rigid translation. 
This is confirmed by the fact that, in the framework of the proposed Ritz approach, the contribution of the rigid displacements to the terms of the energy is  
\begin{subequations}    \label{Eq_41} 
    \begin{align}
    &U_{0,1}^{d}-W_{rm}^{d}=0  \, ,\\
    &U_{0,1}^{h}-W_{rm}^h= 0  \, ,\\   
    &U_{0,1}^{c}-W_{rm}^{c}=0+\frac{1}{2} \pi  q \left(\alpha_1^2+\alpha_2^2\right)=\frac{1}{2} \pi  q \left(\alpha_1^2+\alpha_2^2\right) \,,\\   
    &U_{0,1}^{is}-W_{rm}^{is}=0-\frac{1}{2} \pi  q \left(\alpha_1^2+\alpha_2^2\right)=-\frac{1}{2} \pi  q \left(\alpha_1^2+\alpha_2^2\right) \,,
    \end{align}
\end{subequations}

\noindent and the contribution of the rigid rotation is
\begin{subequations}    \label{Eq_41b} 
    \begin{align}
    &U_{0,1}^{d}-W_{rm}^{d}=- 2 \pi  \beta^2 q R^2-0=- 2\pi  \beta^2 q R^2 \, ,\\  
    &U_{0,1}^{h}-W_{rm}^{h}=- 2 \pi  \beta^2 q R^2+ 2\pi  \beta^2 q R^2=0 \, 
      \, ,\\ 
    &U_{0,1}^{c}-W_{rm}^{c}=- 2\pi  \beta^2 q R^2+ 2 \pi  \beta^2 q R^2=0 \, 
     \,,\\ 
    &U_{0,1}^{is}-W_{rm}^{is}=- 2\pi  \beta^2 q R^2+ 2\pi  \beta^2 q R^2=0 \, 
     \,.
    \end{align}
\end{subequations}

Should one be willing to take into account the generality of the rigid-body motions in the form 
\begin{subequations}    \label{Eq_42} 
    \begin{align}
    & v=C \left(\cos{2\theta}+\alpha_1 \cos\theta +  \alpha_2 \sin\theta  + R\, \beta \right)  \, ,\\
    & w=C \left(2 \sin{2\theta}-\alpha_1 \sin\theta+  \alpha_2 \cos\theta + R \frac{\beta^2}{2}\right)
     \,,  
    \end{align}
\end{subequations}  
the proposed Ritz approach would provide critical multipliers that can be written as 
\begin{equation}  
\begin{split}   
\textit{d.} \quad 
&\lambda=4\left(1+\frac{2}{9} \beta ^2 R^2\right)^{-1}\,,  \\
\textit{h.} \quad 
&\lambda=3\,, \\
\textit{c.} \quad  
&\lambda=\frac{9}{2}\left(1-\frac{\alpha _1^2+\alpha _2^2}{8} \right)^{-1}\,, \\
\textit{is.} \quad  
&\lambda=\frac{9}{4}\left(1+\frac{\alpha _1^2+\alpha _2^2}{16} \right)^{-1}\,.
\end{split}\label{Eq_19tris}
\end{equation}

Formulae~\eqref{Eq_19tris} in some instances may represent an approximate solution of the problem and it is worth noticing that they are able to return, among the others,  the value provided by Chwalla and Kolbrunner in~\cite{chwalla1938beitrage}, i.e. \(\lambda^d=3.265\), and discussed by Singer and Babcock in~\cite{singer1970buckling, singer1971erratum} by setting \(\beta={R}^{-1}\).

The point is that, as said before, the differential equations of the problem have been derived by a variational formulation that has been set up under precise hypotheses on the behaviour of the loads. In this respect, formulae~\eqref{Eq_19tris} are coherent with the solutions~\eqref{EqN_19N} only in the case of \(\beta=0\) for dead load and \(\alpha_1=\alpha_2=0\) for centrally directed loads, representing, conversely, only an energy estimate for the buckling loads~\cite{el1976influence}. 

Also by imposing the continuity conditions~\eqref{Eq_continuity} and the \(2\pi\)-periodicity on the solutions of the differential equations and ruling out any rigid motion, one is left only with those solutions that cannot contemplate any discontinuity provided by the presence of external constraints (as it is the case, for example, of \(\lambda=0.701\), i.e. the one-clamp ring illustrated in Figure 6 of~\cite{gaibotti2024effects}, that it has been obtained by making reference to a differential formulation under specific boundary conditions).

\begin{figure}[h]
    \centering
    \includegraphics[width=\linewidth]{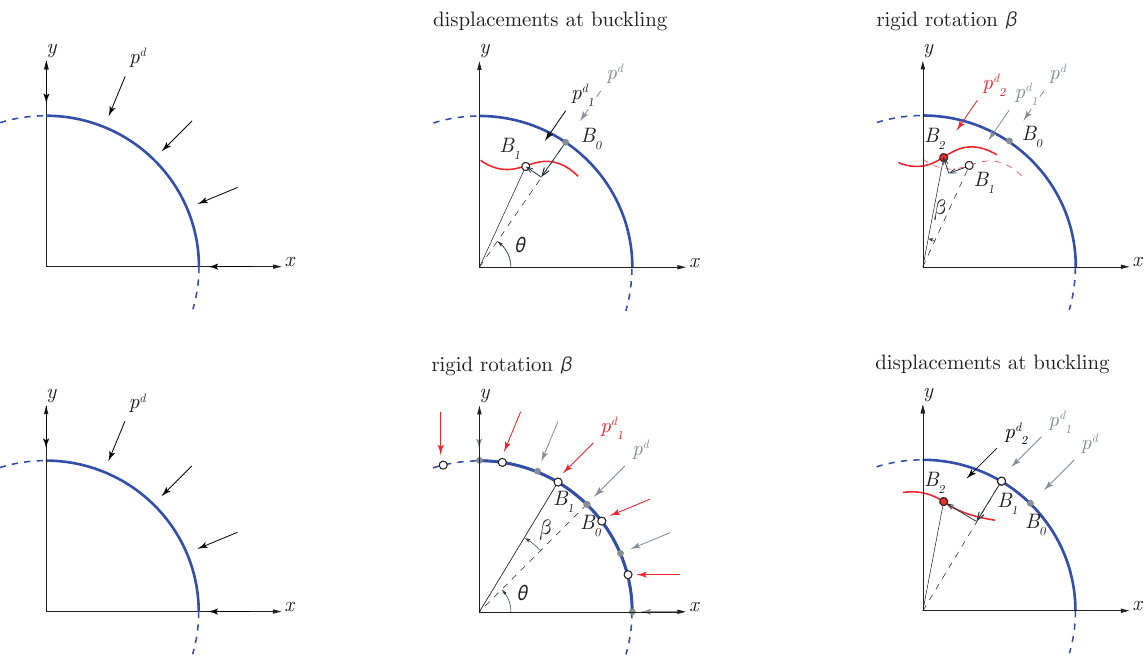}
    \caption{Different displacement paths under dead load, $p^{d}$, in the presence of a rigid rotation \({\beta}\): the ring may first buckle under the load \({p^{d}}_1\) and then rigidly rotate by \({\beta}\) under the load \({p^{d}}_2\) (top panel), or vice versa, it may first rigidly rotate by \({\beta}\) under the load \({p^{d}}_1\) without altering its geometry and then buckle under the load \({p^{d}}_2\) (bottom panel).}
    \label{fig:Fig_5N}
\end{figure}

Actually, as shown in Figure~\ref{fig:Fig_5N}, differently from what happens in the case of hydrostatic pressure, \(p^h\), where the load setup remains the same regardless of the deformation path, in the case of dead load, \(p^d\), the work done by the load on account of any rigid rotation is path-dependent, and the loading condition cannot be considered conservative at all.
In fact, if the ring first buckles and then rotates, the buckling coefficient remains unaffected; conversely, if the deformation path begins with a rigid rotation, the loading arrangement changes, leading to a different setup of the load around the ring, which translates into a different setup of the buckling problem.

\begin{figure}[h]
    \centering
    \includegraphics[width=\linewidth]{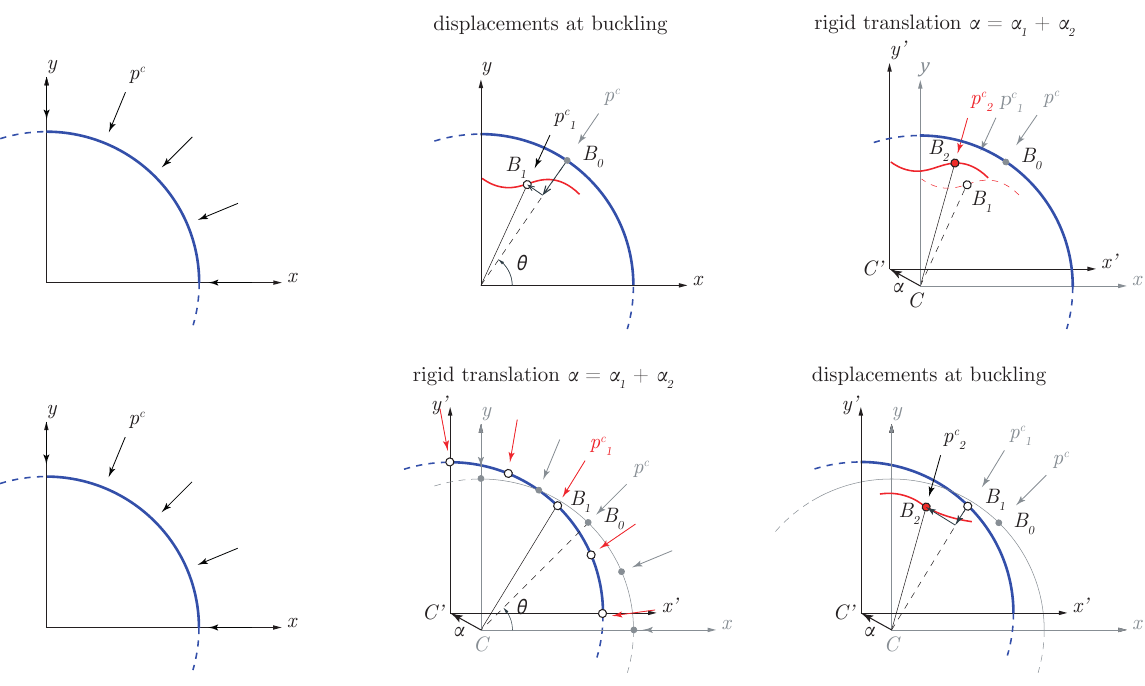}
    \caption{Different displacement paths under centrally-directed load, $p^{c}$, in the presence of a rigid translation \textbf{\(\alpha\)}: the ring may first buckle under the load \({p^{d}}_1\) and then rigidly translate by \({\alpha}\) under the load \({p^{d}}_2\) (top panel), or vice versa, it may first rigidly translate by \({\alpha}\) without altering its geometry under the load \({p^{d}}_1\) and then buckle under the load \({p^{d}}_2\) (bottom panel).}
    \label{fig:Fig_6N}
\end{figure}

The same occurs in the case of a centrally-directed load, \(p^c\), or \(p^{is}\), in the presence of any rigid translation, which induces a change in the position of the ring centre, resulting again in a different definition of the work of the applied load and ultimately leads to a change in the buckling multiplier (see Figure~\ref{fig:Fig_6N}).

\section{Conclusions}

In the present work one of cornerstones of elastic stability, that is, the elastic buckling of circular rings under external pressure, has been re-examined from its roots in the framework of an energy approach and starting from a general non-linear kinematics.
By overhauling the generally accepted paradigm of axial inextensibility and by presenting a solution that fully accounts instead for the pre-buckling extensibility of the middle line of the ring, it has been shown that:
\begin{enumerate}
    \item the classic a-priori inextensibility paradigm cannot straightforwardly lead to the classic solutions in an energy framework where, on the contrary, the extensibility of the ring wall is necessary to allow a unified and meaningful treatment of buckling and initial post-buckling behaviour for a complete variety of cases;
    \item a number of inconsistencies stem from the necessity of adapting the energy formulation to the a-priori assumption of inextensibility, such as the paradox of a zero value of the work of the applied pressure at buckling in the case of dead load and kinematic contradictions involving the first- and second-order terms of the axial deformation of the axis of the ring walls, as well as its curvature;
    \item the pre-buckling deformation of a compressed ring has a significant influence on its post-buckling behaviour, making it stable also in cases when it has been traditionally considered unstable;
    \item the possible influence of rigid motions must be read in the framework of the correct loading hypotheses and it is null only in those cases where the acting pressure can be considered conservative as a whole, a fact that contributes to clarify some issues in literature.
\end{enumerate}

Overall, the proposed unified treatment of pre-buckling, buckling and initial post-buckling of the compressed annular rod marks the fact that its pre-buckling configuration has a much greater influence on its behaviour with respect to what happens in the case of an initially ideal straight column and, differently from Euler’s strut, and, over the years, has led at least to some misunderstandings and misleading results.

\appendix

\section{Energy balance and the issues deriving from the a-priori inextensibility assumption}\label{Appendix C}

The neglectibility of the axial deformation, \(H\lambda\), of the ring with respect to the unity in formulae \eqref{Eq_32tot} is a crucial point because it leads to buckling shapes, which can be defined as quasi-inextensible.

As a matter of fact, most of the published literature over the years (see~\ref{Appendix A}) assumes a-priori that the ring is inextensible.
However, it is immediate to realise that, in such a case, the straightforward application of the Rayleigh-Ritz procedure shown in Section~\ref{Sec: 3} with reference to the expression of the total potential energy~\eqref{Eq_11a} has been used to obtain the Euler-Lagrange equations of the problem, returns incorrect values for the buckling loads~\cite{mascolo2023revisitation}, i.e. no solution in the case \(d.\), \(\lambda=12\) in the case \(h.\), \(\lambda=-36\) in the case \(c.\) and \(\lambda=\nicefrac{36}{7}\) in the case \(is.\) 

The reason for this apparent paradox is not very obvious and deserves a detailed investigation. 

The starting point is that, even if many works point out condition~\eqref{Eq_inex} for inextensibility, in reality they build upon the assumptions of small middle surface strains and moderate rotations, explicitly formalized by Sanders~\cite{sanders1963nonlinear}, a fact that leads to the following assumptions 
\begin{align}
         \varphi=\frac{v-w^{\prime}}{R}=o(\delta) \\
         \varepsilon_{lin}=\frac{w+v^{\prime}}{R}=o(\delta^2)\label{opicc}
\end{align}
where \(\delta<<1\) is a small quantity compared to  unity. In this framework, the inextensibility condition based on equation~\eqref{Eq_1bisa}, that is 

\begin{equation}
    \label{inextensibility}
\varepsilon =\frac{w+v^{\prime}}{R}+\frac{\left(w+v^{\prime}\right)^2+\left(w^{\prime}-v\right)^2}{2 R^2}=0\,, 
\end{equation}
since
\begin{equation}\label{sOapprox}
    \frac{\left({w+v^{\prime}}\right)^2}{2R^2}=o(\delta^4)\approx0
\end{equation}

\noindent is transformed into~\cite{rehfield1972initial,BrushDonOrr1975Bobp,Simitses1976, ThompsonHunt1973} 
\begin{equation}\label{epsred}
     \varepsilon =\frac{w+v^{\prime}}{R}+\frac{\left(v-w^{\prime}\right)^2}{2 R^2}=0\,.
\end{equation}

The work of applied loads~\eqref{Eq_9tot} on account of equations~\eqref{sOapprox} and~\eqref{epsred} thus becomes
\begin{subequations} 
\begin{align}
         \overline{W}_{1}^{d}&=-\frac{p}{2}\int _{0}^{2\pi}\;2Rv^\prime-(v-w^\prime)^2+w\left(w+v^\prime\right)\,d\theta=\nonumber\\
        &=\frac{p}{2}\int _{0}^{2\pi}\;(v-w^\prime)^2-w\left(w+v^\prime\right)\,d\theta
        \;,\label{Eq_9cred}\\
        \overline{W}_{1}^{h}&=
      \overline{W}_{1}^{d}-\frac{p}{2}\int _{0}^{2\pi}\; v\left(v-{w^\prime}\right)\,d\theta \,, \label{Eq_9bred}\\
                \overline{W}_{1}^{c}&=            \overline{W}_{1}^{d}- \frac{p}{2}\int _{0}^{2\pi}\;v^2\,d\theta \;,\label{Eq_9d}\\
               \overline{W}_{1}^{is}&
               =\overline{W}_{1}^{d}- \frac{p}{2}\int _{0}^{2\pi}\;\,d\theta \;,\label{Eq_9e}
\end{align}\label{Eq_9tot1}
\end{subequations}
\noindent and the expression of the total potential energy \eqref{Eq_11a}, i.e.: 
\begin{equation}
{\Pi}^i={U}_0+{U}_1+{U}_{0,1}-(W_0+W_1^i) \;,
\label{Eq_EPT ext}
\end{equation}
\noindent in absence of the extensional terms \(W_0\), \(U_0\) and \(U_{0,1}\), reduces to
\begin{equation}
{\Pi}^i={U}_1 -\overline{W}_1^i \;.
\label{Eq_EPT inex}
\end{equation}

By inserting in the total potential energy~\eqref{Eq_EPT ext} the eigenmodes~\eqref{Eq_19bis}, and making the total potential energy stationary, i.e., \(\nicefrac{\delta\Pi}{\delta C}=0\), the buckling coefficients \(\lambda\) can be readily calculated and coincide with the correct values~\eqref{Eq_19bis}.

However, even if this approach may appear at first sight legit, the devil is in the details, and it is now shown that it underlies an inconsistency in the kinematics of the model.

Actually, in the a-priori inextensible formulation, the non-occurring terms \(U_0\) and \(W_0\) (equations~\eqref{eq_3a} and~\eqref{eq_7N}) are in any case immaterial with respect to the variation \(\nicefrac{\delta U_0}{\delta C}\) and they can be omitted without any harm. 

On the contrary, the mixed term \(U_{0,1}\) (equation~\eqref{eq_3c}) cannot be overlooked and plays a hidden but crucial role, as it will become evident in what follows.   

In fact, the mixed term \(U_{0,1}\), that in the proposed extensible formulation at the onset of buckling reduces to

\begin{equation}\label{eq:U01}
      U_{0,1}=-p R\int_{0}^{2\pi}\frac{\left(w^\prime-v\right)^2}{2 R^2} \, R d\theta= -\frac{p}{2} \int_{0}^{2\pi}    {\left(w^\prime-v\right)^2}    \, d\theta\,,
\end{equation}

\noindent corresponds exactly to the difference \(W_1-\overline{W}_1\), as deducible from Table~\ref{Tab_B1}.
\begin{table}[ht!]
\caption{\(v[\theta]=C\, \cos{2\theta}\) and \(w[\theta]=2C\, \sin{2\theta}\)}
  \centering
  \begin{tabular}{ccccccc} 
  \hline
  \multicolumn{7}{c}{present extensible formulation \eqref{Eq_EPT ext}} \\ 
    Load type  &$U_{0}~^{(1)}$&$W_{0}~^{(2)}$&$U_{1}$ &$U_{0,1}$ &  $W_1-U_{0,1}$&$\lambda$\\ 
    \hline 
    \makecell[c]{\textit{d}. \\\textit{h}. \\ \textit{c}.\\ \textit{is}.}
    &\makecell[c]{$\frac{\pi R^3p_k^2}{EA}$} 
    &\makecell[c]{$\frac{2 \pi R^3p_k^2}{EA}$} 
    &\makecell[c]{$\frac{18 \pi EI}{ R^3}\, C^2$} 
    &\makecell[c]{$-\frac{9 \pi}{2} \,C^2  p_k$} 
    & \makecell[c]{$\frac{9 \pi}{2} \,C^2 p_k$\\ $6 \,C^2 p_k$\\ $ 4\pi\,C^2  p_k$\\ $ 8\pi\,C^2  p_k$}
    & \makecell[c]{$4$\\$3$\\$\nicefrac{9}{2}$\\$\nicefrac{9}{4}$}
    \\ 
    \midrule
    \multicolumn{7}{c}{a-priori inextensible formulation \eqref{Eq_EPT inex}}  \\ 
    Load type   &$U_0$  &$W_0$ &$U_1$ &$U_{0,1}$& $\overline{W}_1$   &$\lambda$\\ 
    \hline 
    \makecell[c]{\textit{d}.\\ \textit{h}.\\ \textit{c}.\\ \textit{is}.}
    &\makecell[c]{$0$}
    &\makecell[c]{$0$}
    &\makecell[c]{$\frac{18 \pi EI}{ R^3}\, C^2$} 
     &\makecell[c]{$0$}
     & \makecell[c]{$\frac{9 \pi}{2} \,C^2 p_k$\\ $6 \,C^2 p_k$\\ $ 4\pi\,C^2  p_k$\\ $ 8\pi\,C^2  p_k$}
        & \makecell[c]{$4$\\$3$\\$\nicefrac{9}{2}$\\$\nicefrac{9}{4}$}\\
    \bottomrule
  \end{tabular}\label{Tab_B1}
\end{table}

As a matter of fact, in the a-priori inextensible formulation, the ignored mixed term, \(U_{0,1}\), re-emerges — almost as if by magic — as a contribution to the work of applied loads. 
The main drawback to this approach is that in doing so the extensibility of the ring is not accounted for in the pre- and initial post-buckling behaviuor, as it is instead done in the proposed extensible formulation with all the consequent implications for the analysis. More importantly, it hides two kinematic inconsistencies also at the onset of buckling. 

First, expressions~\eqref{Eq_9tot1} are obtained by virtue of equation~\eqref{epsred}, which, however, is not satisfied by the eigenfunctions~\eqref{Eq_19bis}.

Second, since the eigenfunctions~\eqref{Eq_19bis} satisfy the condition \({\varepsilon}_{lin}=0\), should the imposed condition~\eqref{epsred} actually be satisfied, it would imply \((v-w^\prime)^2=0\), that is \(\varphi=0\) and consequently \(\chi=0\)~\cite{babilio2023static}.

In other words, it is evident that any explicit or hidden~\cite{budiansky1974theory} manipulation of the energy formulation in the case of assumed a-priori inextensibility of the ring seems to have been implemented with the precise goal of obtaining the classical buckling results at the price of some kinematic inconsistencies. 

In the present contribution, it has been shown that this is not the case because the actual buckling eigenfunction results naturally quasi-inextensible (in the sense of nullity of~\eqref{Eq_13a}) at the onset of buckling and only at this stage.   

%
%
\section{Starting from Levi: some milestones in the evolution of ring buckling story}\label{Appendix A}

The stability of uniformly compressed rings has a very long history, marked by several landmark contributions to its evolution. Here is a brief overview of the key developments that seem to have led, to the best of the present authors' knowledge, to the current state of the art. 

\begin{enumerate}
    \item[1884. ] 
Levy~\cite{levy1884memoire} can be considered to have provided in 1884 the foundational insights that paved the way for all the subsequent investigations on the topic. Actually, he pioneered the idea that ring instability takes place under conditions of quasi-inextensibility, a circumstance that occurs for hydrostatic load at the classic stability coefficient \(\lambda=3\), he wrote: 

"\emph{[...] D'où je conclus que la limite de \(4/9\) que nous avons trouvée ne peut pas différer de la limite la plus faible possible, de plus de \(4/9\). Au point de vue pratique, il n'y a pas d'inconvénient à prendre h un peu trop fort; au point de vue théorique, il est présumable que la limite la plus faible possible est celle qui répond à la déformation infiniment petite, c'est-à-dire \(3/9=1/3\), parce que, si l'on a donné à un anneau une forme telle qu'il ne puisse pas se déformer infiniment peu, il est extrêmement probable qu'a fortiori il ne pourra pas prendre une déformation finie. Toutefois, si cette présomption est exacte, elle doit pouvoir se déduire rigoureusement des équations (16) qui définissent U et u, et c'est ce à quoi je n'ai pas réussi. La question mériterait donc, au point de vue théorique, d'être complétée en ce sens.}"

From this incipit, the major contribution to the buckling of inextensible\(\slash\)quasi-inextensible rings can be considered the following: 

    \item[1938. ] Chwalla and Kollbrunner~\cite{chwalla1938beitrage} employed an energy approach to the more general instability problem of a circular arch subjected to a uniform inward pressure per unit circumferential length. Their approach distinguishes between two different buckling stages: the first or pre-buckling stage, in which the arch/ring contracts elastically due to axial compression storing potential energy, and the second stage, where the arch/ring can assume a buckled non-circular configuration. However, at the second stage they impose the a-priori kinematic condition~\eqref{Eq_inex} that, as discussed in the present work, prevents a correct formulation of the energy problem.

    They found the following critical coefficients
    
     \begin{itemize}
        \item[\textit{h}.]\quad \(\lambda=3\)\,,
        \item[\textit{d}.]\quad \(\lambda=3.265\)\,,
        \item[\textit{c}.]\quad \(\lambda=4.5\)\,.
    \end{itemize}

    where the case of dead loading, \(d\), results incorrect.
    
    \item[1952.] In order to derive the critical pressure of a thin circular ring under hydrostatic load, Timoshenko and Gere~\cite{timoshenko1985theory} made resort to a direct and very simple geometrical approximate approach, assuming small deflections and inextensional deformations along the centreline of the ring. More in detail, their analysis disregards the displacements in the tangential direction, \(v\), and derives the bending moment from the approximate  curvature of an element after deformation. 
    
    The following second-order differential equation\footnote{Notation according to~\cite{timoshenko1985theory}: \(w\): inward radial component of displacement; \(\theta\): angular coordinate of the deformed ring; \(R\): radius of the undeformed circular ring; \(EI\): flexural rigidity, \(M\): bending moment.} is thus obtained 
     
    \begin{equation}
        \label{w_Timos}
        \frac{d^2w}{d\theta^2}+w=-\frac{M\, R^2}{EI}\,,
    \end{equation}

    which is then used to derive the classic solution \(\lambda=3\) for hydrostatic load. 
    It is interesting to note that equation \eqref{w_Timos} is the same equation found by Biezeno and Grammel (cf.~\cite{Biezeno1953}, Equation (4), page 340) on the basis of a slightly different geometrical approach.

    \item[1952.]  
    Stevens~\cite{stevens1952stability} made a critique of the approach of Timoshenko, stating that

    "\emph{S. Timoshenko bases his treatment upon}" equation \eqref{w_Timos} 
    "\emph{though}"  [...] 
    "\emph{he has not used it correctly. This particular equation must be irrelevant in the initial stages of the proof. His approach leads to a second-order differential equation in \(w\) for the stability condition for a ring under constant pressure. No clear lead is given of the steps to generalize to the case where the pressure varies with displacement. The logical method, however, is to examine the equilibrium of an element of a ring subjected to the components of force and to the turning moments.}"
    On the contrary, Stevens, starting from the same equilibrium equation provided by Biezeno and Koch~\cite{Biezeno1945Koch}\footnote{Notation according to~\cite{stevens1952stability}, that is \(s\): the hoop distance along the circumference; \(c\): radius of curvature of the element; \(EI\): flexural rigidity; \(C\):  compression force in the ring; \(M\): bending moment, \(t\): distributed load per unit length tangential to an element of the ring},
    \begin{equation}
        \label{Biezeno5Koch}
        \left(\frac{1}{c}-\frac{M}{EI}\right)\frac{dM}{ds}+\frac{dC}{ds}=t\,,
    \end{equation}
 \noindent and following an integral approach, derives a generalized stability criterion for a ring subjected to radial loads, which fully accounts for both shear and compressive forces but, under the assumption of a thin ring, again neglects both shear and axial deformations.

    When a generic load remains normal to the ring after deformation and the tangential force can be neglected as a second-order effect, which mimics the case of a hydrostatic load, Stevens erroneously found that for small deformation after the occurrence of buckling the circular ring becomes unstable. 
  
    \item[1957.] Bodner~\cite{bodner1958conservativeness} focused on the conservativeness of different systems of force generally considered in the classic treatment of the ring buckling and provided the first thoughtful  discussion of the matter, together with some interesting considerations. He reported the classic correct values for the critical load of a ring.
    
    With respect to the hydrostatic pressure, \textit{h}., he stressed that the proof of its conservativeness can be found in text on hydrodynamics (e.g., pp. $13$ and $14$ of reference~\cite{Milne1952}). 
    
    In the case of dead pressure, he states that "\emph{Since the work done by a force that is constant in direction and magnitude depends only upon the initial and final configurations, this system is conservative. Although such a force field is easily defined, it would be difficult to devise an experimental setup in which such a field is applied to a ring.} 

    As pointed out in the present work this is true in absence of any rigid rotation.
    It is concluded that, similarly to what happens in the Euler column, buckling in that case "\emph{is governed only by the energy of the induced bending stresses and the part of the work done by the pre-buckling membrane stresses that involves non-linear displacement terms}".

    Finally, the conservativeness of centrally directed force is justified by noting that "\emph{the work done by such a unit force during buckling is the product of the force and the change in the distance of the element from the pole. The work is therefore independent of the path so that the force field is conservative. If the force intensity were not constant but depended in a continuous and single-valued manner upon the displacement from the pole, then the system would still be conservative, e.g., the general central force problem.}" 

    \item[1956.] Wasserman~\cite{wasserman1961effect} investigated the free in-plane and three-dimensional vibrations of a ring for the hydrostatic, dead and centrally directed loads. In the hypotheses of negligible transverse force effect, negligible inertia of rotation and inextensibility of the ring centreline, he obtains the equations for the vibration in terms of radial displacement.

    By reducing the free vibrations to zero, the dimensionless load parameter \(\lambda\) for which the load attains the classic critical value for all three considered load cases. 

    Wasserman, as previously Bodner, provides also a physical interpretation of the active and restoring forces competing in the buckling phenomenon.

    \item [1969.] Oran et al.~\cite{oran1967complementary} studied the effect of load behaviour on the linear buckling of uniformly compressed circular arches. 
    The buckling behaviour is derived by imposing a priori the condition of inextensible deformations in the form
    ~\eqref{Eq_inex} and gives origin to a sixth-order differential equation whose general solution is determined by imposing different types of symmetric boundary conditions that ensure the conservativeness as a whole of the loading system. 
    
    On the conservative character of configuration-dependent lateral pressure, Oran states that
    
    \emph{When the external loads are assumed to rotate with the axis of the arch, the problem has the important feature of being "locally nonconservative". In the case of uniformly compressed arch, however, the problem becomes conservative "as a whole", when certain geometric restrictions (to be examined subsequently herein) are imposed on the translations of the end sections of the arch.}

    Oran et al. mean by \emph{locally conservative} that the  work done by the load on an isolated infinitesimal arch element depends on the specific deformation path. Moreover, he stressed that
    
    \emph{The writers' decision to present their study within the context of conservative versus non conservative problems stemmed from their feeling  that the conservative character of normal loads, although conditional, is often taken for granted. The misconception appears to be due to the fact that a uniformly distributed normal load is generally visualized as a hydrostatic pressure, the conservative nature of which readily follows from physical considerations. However, for this representation to be valid, the (presumably equivalent) hydrostatic loading must be "physically realizable"} [...]. \emph{If the bar is only loaded over a portion of its length, or if one end does not follow a prescribed path (as in the case of a cantilever bar) no equivalent hydrostatic loading can be physically realized; therefore, the problem becomes nonconservative.}

    Actually, Oran et al. add further considerations on the conservativeness of the loading.
    
    \item[1970.] Singer and Babcock~\cite{singer1970buckling} limit the analysis of the problem to small displacements of an a-priori inextensional ring but provide the first comprehensive, interesting and in-depth discussion on the effect of the rigid-body displacements on the buckling pressure, which leads them to conclude that the buckling coefficient \(\lambda=3\) for the hydrostatic pressure and \(\lambda=4.5\) usually reported in literature, are correct. But the value \(\lambda=4\) for the dead load is "\emph{meaningless without specification of constraints preventing a rigid-body rotation of the ring}". They state that the value \( \lambda=3.265\), (later corrected to \( \lambda=3.273\) by Singer and Babcock~\cite{singer1971erratum}),  that was first reported by Chwall and Kollbrunner~\cite{chwalla1938beitrage} is correct when the ring is fixed in space at two diametrically opposite points, while the value \( \lambda=2.25\) reported by Stevens~\cite{stevens1952stability} for the centrally directed pressure is due to a sign error and then incorrect.
    \item[1972.] Rehfield~\cite{rehfield1972initial} is one of the first authors to discuss the postbuckling behaviour of the ring under hydrostatic and uniform radial loadings. The approach is based on asymptotically power expansions of the amplitude of the buckling mode, according to Koiter's approach. His study suggests that both types of loading exhibit a symmetric point of bifurcation, with the ring under hydrostatic pressure being slightly imperfection-sensitive, with an unstable post-buckling behaviour. The ring under uniform radial loads, instead, is neutrally stable to the first approximation. 
    Such erroneous conclusions are probably due to inadequate formulations, as pointed out by El Naschie in~\cite{el1976influence}.
    
    \item[1975.] In the years between \(1974\) and \(1976\) El Naschie studied ring buckling and postbuckling extensively~\cite{el1976influence,el1975conservativeness}, emphasizing repeatedly the importance of loading behaviour and of supporting conditions.
    
    In particular, he analysed, always under the hypothesis of a-priori inextensibility, a ring sliding freely on the horizontal diameter and a ring sliding freely on both horizontal and vertical diameters.

    \item[1980.] Schmidt~\cite{schmidt1980critical} contributed to the debate in the framework of the general non-linear theory of curved beams and in the hypothesis that the slender number \(\frac{I}{R^2 A}\) and the extensional centroidal strain are negligible quantities with respect to the unity, as well as the deflections during buckling and obtains for a symmetric buckling mode under dead load the classical critical values
    \(\lambda=4,\,9,\,16,\dots\).
    In the case of an asymmetric mode of buckling of an hingeless arch with its immovably clamped ends, he finds that the critical load is considerably lower than that obtained for the symmetric mode, that is 
    \(\lambda=0.701\). However, this value does not represent any solution for the characteristic equations derived for a whole ring under the conditions~\eqref{Eq_continuity}. 
    
    Importantly, he states that
    "\emph{An experimental verification of the foregoing results, although not easy, is possible. For that purpose, we could take a smooth horizontal table and drill through it a large number of closely spaced, equidistant holes of small diameter along a large circle of radius b. Next, we place, concentrically with the circle of holes, a thin circular ring (or arch) of radius a on this table, attach radial strings to the ring and pass them through the drilled holes in such a way as to be able to produce a fairly uniform compression in the ring by attaching weights to the strings' ends.}"
    Such an experimental setup has been very recently made by Gaibotti et al.~\cite{gaibotti2024effects}.

    \item[1973.] In their treatise, Thompson and Hunt~\cite{ThompsonHunt1973} focused on the critical dead pressure. They indicate a coefficient of stability equal to $4$ and express the buckling mode as $w(\theta)= a \, \cos{2\theta}$ and  $v(\theta)= \nicefrac{1}{2}a \, \sin{2\theta}$. They also put emphasis on the fact that     

    "\emph{The Donnell equations and the improvement considered here are appropriate for problems in which the middle-surface membrane stresses have a strong influence on the initial post-buckling behaviour [\dots] membrane action is absent in the behaviour of struts and rings and the use of the Donnell-type equations will therefore predict zero or near zero post-buckling curvature. For such problems there will of course be a rather weak post-buckling contribution, due to non-linear bending action and therefore dependent on EI, but this could not be picked up by our proposed extensible formulations since it requires an improvement on the x-w relationship corresponding to that of the Euler strut analysis.}"
     Actually, they hint at some of the issues which have been extensively discussed in the present study in Section~\ref{Sec: 2}
    \end{enumerate}

Among the authors who tried to account somehow for the extensibility of the axis of the circular rod, it is worth mentioning the following: 

\begin{enumerate}
  \item [1955.] Boresi~\cite{boresi1955refinement} pursued  the buckling pressure by means of a numerical energetic approach. In the expression of the strain components, the second-degree terms of the initial radial displacements are neglected. Due to the complexity of an exact treatment of the problem, he contented himself with providing a lower and an upper bound for the buckling pressure as functions of the thickness-radius and Poisson ratios for hydrostatic and centrally directed loads.  The  lower and upper bounds converge as the thickness of the ring approaches zero. 
  
  For a vanishing thickness to radius ratio and quasi-negligible Poisson ratio, he attained once more the classical values of the buckling pressures.

    \item [1956.] Pearson~\cite{pearson1956general} focused on the effects of non-linearity and of the load behaviour by comparing an adjacent-equilibrium position by energy techniques. He found that these approaches are quite equivalent and that

    \begin{itemize}
        \item[\textit{h}.]\quad \(\lambda=\frac{3}{(1-\nu^2)}\)
        \item[\textit{d}.]\quad \(\lambda=\frac{4}{(1-\nu^2)}\)\,,
    \end{itemize}
thus providing an explicit dependence on the Poisson's ratio for thick rings.
    
   \item[1969.] Smith and Simitses~\cite{smith1969effect} derived the critical pressure for the three classical load behaviours in the framework of the classical ring theory, including the effect of transverse shear and not invoking the inextensibility assumption. The equilibrium equations and the stress-strain relations are obtained through the principle of virtual work and the bifurcation points by means of the adjacent equilibrium criterion. By ignoring the transverse shear deformation, that is, by letting the shear coefficient be zero \(k_s\) in the following relationships, their results are in full agreement with the classical one 

   \begin{itemize}
        \item[\textit{h}.]\quad \(\lambda=\frac{3}{(1+4 k_s)}\)\,,
        \item[\textit{d}.]\quad \(\lambda=\frac{4}{(1+4 k_s)}\)\,,
        \item[\textit{c}.]\quad \(\lambda=\frac{4.5}{(1+4 k_s)}\)\,.
    \end{itemize}
    
    With respect to rigid motions, they state: "\emph{The reason for considering the three different cases is because all those have
    been used as models for the real load case, which is pressure. The pressure behavior is best represented by case I. It is difficult to conceive of a true application that is represented by case II. Case III can serve as a mathematical model for the following problem: Consider a thin ring that is loaded by a very large number of closely spaced radial cables pulled together through a stiff, small, rigid ring at the center of the thin ring. Singer and Babcock (1970) have shown that, for case II, the thin ring is unstable as a rigid body and will rotate under arbitrarily small pressure.}"

    \item [1967.] By assuming that the centreline of the ring is not extensible, K{\"a}mmel~\cite{kammel1967einfluss} found the critical pressure in the case of hydrostatic, and dead load to be functions of the powers of the thickness-radius ratio as
    
   \begin{itemize}
        \item[\textit{h}.]\quad \(\lambda=3\left(1+\frac{8}{5}\left(\frac{h}{R}\right)^2+\frac{416}{105}\left(\frac{h}{R}\right)^4+\dots\right)\)\,,
        \item[\textit{d}.]\quad \(\lambda=4\left(1+\frac{8}{5}\left(\frac{h}{R}\right)^2+\frac{127}{35}\left(\frac{h}{R}\right)^4+\dots\right)\)\,.
    \end{itemize}

     \item[1975.] Finally, for an extensional ring, El Naschie~\cite{el1975extensible} proposed an expression for the pressure-deformation relationship of a ring under uniform constant directional pressure.
\end{enumerate}

 \bibliographystyle{elsarticle-num} 
 \bibliography{cas-refs}





\end{document}